\documentclass[aps,prl,reprint,superscriptaddress]{revtex4-1}

\usepackage{amsmath}
\usepackage{graphicx}
\usepackage{color,longtable}
\usepackage{epstopdf}

\newcommand{\expect}[1]{\langle {#1} \rangle}
\newcommand{\br}{{\bf r}}
\newcommand{\bp}{{\bf p}}
\newcommand{\ph}{{\hat \psi}}

\begin{document}

\title{Spinor dynamics in an antiferromagnetic spin-1 thermal Bose gas }

\author{H. K. Pechkis}
\affiliation{Quantum Measurement Division, National Institute of Standards and Technology, and Joint Quantum Institute, NIST and University of Maryland, 100 Bureau Drive, Gaithersburg, MD, 20899-8424, USA }
\author{J. P. Wrubel}
\affiliation{Quantum Measurement Division, National Institute of Standards and Technology, and Joint Quantum Institute, NIST and University of Maryland, 100 Bureau Drive, Gaithersburg, MD, 20899-8424, USA }
\affiliation{Physics Department, Creighton University, Omaha, NE, 68178, USA}
\author{A. Schwettmann}
\affiliation{Quantum Measurement Division, National Institute of Standards and Technology, and Joint Quantum Institute, NIST and University of Maryland, 100 Bureau Drive, Gaithersburg, MD, 20899-8424, USA }
\author{P. F. Griffin}
\affiliation{Quantum Measurement Division, National Institute of Standards and Technology, and Joint Quantum Institute, NIST and University of Maryland, 100 Bureau Drive, Gaithersburg, MD, 20899-8424, USA }
\affiliation{Department of Physics, University of Strathclyde, Glasgow G4 0NG, UK}
\author{R. Barnett}
\affiliation{Department of Mathematics, Imperial College London, London SW7 2AZ, UK}
\affiliation{Joint Quantum Institute and Condensed Matter Theory Center, Department of Physics, University of Maryland, College Park, Maryland 20742, USA}
\author{E. Tiesinga}
\affiliation{Quantum Measurement Division, National Institute of Standards and Technology, and Joint Quantum Institute, NIST and University of Maryland, 100 Bureau Drive, Gaithersburg, MD, 20899-8424, USA }
\author{P. D. Lett }
\affiliation{Quantum Measurement Division, National Institute of Standards and Technology, and Joint Quantum Institute, NIST and University of Maryland, 100 Bureau Drive, Gaithersburg, MD, 20899-8424, USA }

\date{\today}

\begin{abstract}
We present experimental observations of coherent spin-population oscillations in a 
cold thermal, Bose gas of spin-1 $^{23}$Na atoms. 
The population oscillations in a multi-spatial-mode thermal gas have the same behavior as those observed in a single-spatial-mode antiferromagnetic spinor Bose Einstein condensate. We demonstrate this by showing that the two situations are described by the same dynamical equations, with a factor of two change in the spin-dependent interaction coefficient, which results from the change to particles with distinguishable momentum states in the thermal gas.   We compare this theory to the measured spin population evolution 
after times up to a few hundreds of ms, 
finding quantitative agreement with the amplitude and period.  We also measure the damping time of the oscillations as a function of magnetic field. 
\end{abstract}

\pacs{67.85.-d, 05.30.Jp, 51.10.+y, 51.60.+a}

\maketitle

Although Bose-Einstein condensates (BECs) are often thought of for sensitive measurements, their spatial coherence is not always necessary.  
Thermal atomic collisions are often mistakenly thought to be incoherent but, 
while keeping track of the spatial coherence is difficult, coherence can sometimes more easily be followed in the internal degrees of freedom.
Thus, cold thermal clouds are often just as sensitive for use in spin measurements.  In this work, we 
demonstrate collisionally-driven coherent spin population oscillations that can be interpreted as zero-momentum spin waves in a cold thermal cloud of spin-1 atoms. 
Such oscillations were previously only seen in the context of BECs \cite{Chang, Black, Kronjager, Kronjager2, Higbie} and two-atom, single-spatial-mode systems \cite{Widera, Widera2}. 
The spin oscillations that we observe in a highly multi-spatial-mode thermal gas are remarkable in that they can be described by a theory that is independent of the spatial degrees of freedom.
 
Well-known examples of thermal spin systems that preserve internal spin states include optically-pumped dilute gases used for
magnetometry \cite{Brown} and spin-polarized noble gas imaging \cite{Chupp}. The spin polarization can be maintained even while the gas is trapped in glass cells, or by living tissues like lungs. 
Hydrogen masers are based on interrogating the free precession of a spin superposition of a thermal gas in a glass cell.
Less well-known, collisionally-driven spin-wave effects were predicted in 1982 \cite{Lhuillier, Laloe}, and observations of such effects were reported soon thereafter in low-temperature spin-polarized hydrogen \cite{Johnson}.  Bosonic and fermionic alkali pseudo-spin-1/2 systems have also been studied \cite{Oktel, Fuchs, Williams}, and spin domain formation has been observed in these systems \cite{Lewandowski, McGuirk, Du}. 
Due to the spin-dependent interaction that is absent in the pseudo-spin-1/2 system, a spin-1 gas is predicted to have additional interesting coherent collisional (spinor) dynamics which give rise to spin waves or population oscillations \cite{Endo2008, Natu}.

The dynamics of spinor BECs have been widely investigated in spin-1 Na and Rb gases, as reviewed in Refs.~\cite{Kawaguchi, StamperKurn}.  
Rb in the $F=1$ state is ferromagnetic (spin-aligned collisions having the lowest energy), whereas Na is considered antiferromagnetic. Both of these systems display interesting population dynamics in spinor BECs \cite{Chang, Black}. 
Condensed spin systems with $F=2$ and $F=3$ have also been studied \cite{Laburthe, Sengstock02}. 
Single-spatial-mode BEC systems can be modeled analytically \cite{Pu,You} 
and the system can undergo regular oscillations such as those seen in Fig.~1(a).  Figure 1a also shows that when even a few spatial modes are excited in the cloud the regular oscillations break down. In an extended BEC interactions lead to the formation of spin domains \cite{Stenger, Hall, Sadler}.

\begin{figure}[h]
\centering
\includegraphics[width=1\linewidth]{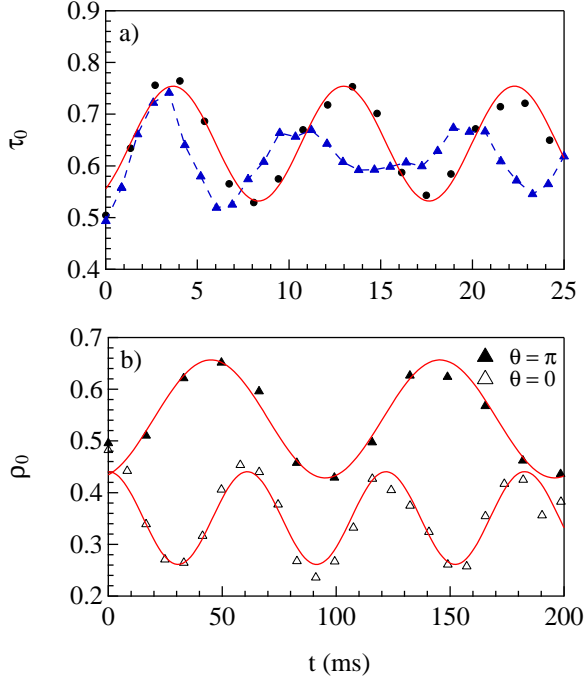}
 \caption{\label{Fig.1} (Color online) (a) Spin population oscillations of a spin-1 Na spinor BEC.  The fractional $m_F=0$ population is shown as a function of time.  The black circles represent a single-mode BEC with initial relative phase $\theta = \pi$ and magnetic field $B = 37 ~\mu$T. Here, $\theta = \theta_1 + \theta_{-1} - 2 \theta_0$, and $\theta_i$ is the phase of spin component i. The blue triangles correspond to a multi-spatial-mode BEC ($\le$5 modes) 
with initial $\theta = \pi$ and $B = 42 ~\mu$T. The solid curve is a sine wave fit to the data.  The dashed line is a guide to the eye. 
(b) Fractional $m_F=0$ spin population of a cloud of thermal spin-1 Na atoms as a function of time with sinusoidal fits shown as solid lines. The solid (open) triangles are for an initial condition of $\theta = \pi$ ($0$) and $B = 11 ~\mu$T ($21 ~\mu$T).
Note the very different time scales.
}
\end{figure}

Here we present experimental observations of coherent spin-population oscillations in a thermal spin-1 Bose gas. Figure~1b shows such coherent oscillations driven by spin-mixing collisions, which are surprisingly similar to those seen in a spinor BEC.
In fact, we find good agreement with predictions from the same dynamical equations as satisfied by a single-mode spinor BEC, even though in a thermal gas many thousands of spatial modes are populated. 
In the BEC, the spin-dependent interaction energy for each added particle is smaller than Planck's constant times the trap frequency, so that the formation of spin domains is suppressed, the spatial degree of freedom stays single-mode, and it factors out of the problem. 
In the thermal gas we have a case where the thermalizing (spin-independent) collision rate is much smaller than the trap oscillation rate, but larger than the spin-dependent collision rate.  This preserves the spatial and velocity distributions and prevents the build up of any correlations between spin and spatial modes.
The average density stays constant in time, and the spatial degree of freedom factors out.

We demonstrate this reduction to a spin-only description by deriving a Boltzmann transport equation under conditions of a time-independent solution in momentum and space and showing that the resulting spin density  satisfies the same equations as for a single spatial mode with a modified spin-dependent interaction coefficient. 
The spin-dependent interaction coefficient $c_2$ must be replaced by 2$c_2$; the change is the distinguishable-particle factor of two that appears for the thermal case \cite{Stoof}.

We start with the second-quantized Hamiltonian describing atoms in a magnetic field $B$ and trapping potential $V(\br)$ \cite{Ho,Ohmi},
\begin{align}
\label{eq:H}
\mathcal{H} = \int d \br& \left[ \ph_a^\dagger \left(
-\frac{\hbar^2}{2m} \nabla^2 + V(\br) + q S_z^2\right) \ph_a
\right. \\
&+  \left.  \frac{c_0}{2} \ph_a^\dagger \ph_b^\dagger   \ph_a  \ph_b  +\frac{c_2}{2} \ph_a^\dagger \ph_c^\dagger {\bf S}_{ab} \cdot
{\bf S}_{cd}\ph_b \ph_d 
\right]\,, \notag
\end{align}
where the $\ph_a(\br)$ are bosonic field operators for spin projection $a$ in the interaction picture that incorporates the linear Zeeman effect.  $m$ is the atomic mass, $\hbar$ is Planck's constant divided by $2\pi$, $q S_z^2$ is the quadratic Zeeman interaction with $q = \gamma B^2$, and $\gamma/h = 27.7$  kHz/(mT)$^2$ for Na. ${\bf S}=(S_x, S_y,S_z)$ is a vector of matrices, where the $S_{\alpha}$ are the spin-one matrices. 
We use the convention of summing over repeated indices. The interaction parameters $c_0$ and $c_2$  can be expressed in terms of the $s$-wave scattering lengths $a_0$ and $a_2$ for two colliding atoms with total molecular angular momentum 0 and 2, i.~e.,~$c_0 = 4\pi \hbar^2 (a_0 + 2 a_2)/(3m)$ and $c_2 = 4\pi \hbar^2 (a_2 - a_0)/(3m)$. 
The Hamiltonian conserves the total atom number and magnetization.  
$^{23}$Na is antiferromagnetic with $c_2>0$.
Only $s$-wave collisions need to be considered here as barrier heights for other partial waves are much higher than our typical thermal energies.  In fact, for a spin-1 system Bose symmetry would also prevent $p$-wave collisions from contributing to spinor oscillations.  Hyperfine-changing collisions to the $F = 2$ state are also not energetically allowed.

A kinetic theory for a thermal gas of spinor bosons with no condensate component can be developed by assuming $\expect{\ph_a} = 0$ for all $a$ and assuming that the spatial and momentum distributions are time-independent.  To this end, we introduce the Wigner density operator \cite{Wigner}
\begin{align}
\label{eq:wigner}
\hat{f}_{ab}(\br,\bp)=\int d\br' e^{-i \bp\cdot \br'/\hbar} \ph_b^\dagger(\br -\br'/2) \ph_a(\br+\br'/2).
\end{align}
A Boltzmann transport equation can be derived by computing the Heisenberg equations of motion of $\hat{f}_{ab}$ using Eq.~(\ref{eq:H}), taking the expectation value with the initial state, defining $f_{ab}(\br,\bp,t) \equiv \expect{e^{i{\cal H} t/\hbar}\hat{f}_{ab}(\br,\bp)e^{-i{\cal H} t/\hbar}}$,
and applying the Hartree-Fock approximation.  Following this procedure, one arrives at the equation
\begin{equation}
\label{Eq:BE}
\partial_t f + \frac{\bp}{m} \cdot \nabla_\br f - \nabla_\br V \cdot
\nabla_\bp f + \frac{i}{\hbar} [f,M] +\frac{1}{2}\{\nabla_\bp f,\nabla_\br
M\}=0,
\end{equation}
where $f$ is the matrix with components $f_{ab}$ and curly and square brackets denote anticommutators and commutators, respectively.  In this equation, we have introduced the position-dependent matrix
\begin{equation}
M(\br)= c_0{\rm Tr }(n) + c_0 n + c_2 {\rm Tr}(S_{\alpha} n) S_{\alpha} + c_2 S_{\alpha} n S_{\alpha} 
+ q S_z^2,
\end{equation}
with the spin position-density matrix 
$n(\br, t)\equiv \int d\bp f(\br,\bp,t)/(2\pi \hbar)^3 $ and the repeated index $\alpha$ is summed over $x,y$, and $z$.
This transport equation was obtained previously in \cite{Endo2008,Natu} and details of its derivation can be found there.  We do not include the collision integral, which has the role of describing damping.

We search for solutions of Eq.~(3), making the ansatz
\begin{equation}
\label{Eq:f}
f_{ab}(\br,\bp,t) = e^{-[p^2/(2m) +V(\br)]/(kT) } \sigma_{ab}(t)/{\cal Z} ,
\end{equation}
under the assumption that the momentum and spatial profiles of the gas have a Boltzmann distribution with temperature $T$, and the dynamics of the spinor degrees of freedom are contained in $\sigma_{ab}(t)$. Here, $k$ is the Boltzmann constant, ${\cal Z}$ is chosen such that ${\rm Tr}(\sigma(t))=1$ and $ \int d\bp d\br \,{\rm Tr}[f(\br,\bp,t)]/(2\pi\hbar)^3 = N$, where $N$ is the total atom number.  Upon inserting Eq.~(\ref{Eq:f}) into Eq.~(\ref{Eq:BE}) and integrating over position and momentum, the equations of motion reduce to
$i \hbar \partial_t \sigma = [\sigma,M_{\rm TG}]$
where
\begin{equation}
\label{Eq:Msm}
M_{\rm TG}=2 c_2 \bar{n} {\rm Tr}(S_{\alpha} n) S_{\alpha} 
+ q S_z^2
\end{equation}
and $\bar{n} = \int d\br [{\rm Tr}(n(\br))]^2/N$ \cite{footie}.

While this formalism can treat mixed-state spin densities, we only prepare pure spin states (although mixed spatial states) in the experiment. Such states remain pure under our equations of motion and the spin density can be written as $\sigma_{ab}=  \sqrt{\rho_a}\exp(-i \theta_a) \sqrt{\rho_b}\exp(i \theta_b)$
, where  $\rho_a$ and  $\theta_a$ are the population and phase of spin component $a$.

For comparison, we consider the mean-field dynamics in a single-spatial-mode BEC without any thermal component \cite{You}.  It is given by 
\begin{equation}
i \hbar \partial_t \phi = c_2 \bar{n} (\phi^\dagger { S_{\alpha}} \phi)  {S_{\alpha}} \phi + q S_z^2 \phi,
\end{equation}
where $\phi=(\phi_1, \phi_0,\phi_{-1})^T$ is the vector order parameter of the condensate, which is normalized to $\sum_a |\phi_a|^2=1$. 
To make a connection with the thermal case, we introduce the matrix $\tau_{ab} = \phi_a^* \phi_b$.  It is straightforward to compute the equation of motion
$
i \hbar \partial_t \tau= [\tau,M_{\rm BEC}]
$
where 
\begin{equation}
\label{Eq:Mbec}
M_{\rm BEC}=c_2 \bar{n} {\rm Tr}(S_{\alpha} n) S_{\alpha} 
+ q S_z^2.
\end{equation}

By comparing with Eq.~(\ref{Eq:Msm}), one sees that the dynamics of the thermal gas can be obtained from the BEC case by simply doubling the spin-interaction coefficient $c_2$. 
Reference \cite{You} shows that, by taking into account the conservation laws, the system can be described by two real variables, $\rho_0$ and $\theta=\theta_{+1}+\theta_{-1}-2\theta_0$.  In addition, they derive expressions for the period and amplitude of these oscillations as a function of $c_2\bar n$, the magnetic field, and the initial state.

Our experimental apparatus consists of a 1070 nm crossed-optical-dipole trap loaded with Na atoms in the $F = 1$ hyperfine ground state.  We prepare approximately $9 \times10^4$ thermal atoms with a temperature of $\approx$ 1 $\mu$K  after a brief period of optical molasses and 6.25 s of forced evaporation \cite{Black}. A weak magnetic field gradient is applied during the evaporation to polarize the atoms into either the $m_F=-1$ or $m_F=0$ sublevel.
After the evaporation, the magnetic field is changed adiabatically to a uniform field that splits magnetic sublevels by 1 MHz. Microwave sweeps are used to selectively transfer any atoms remaining in the unwanted magnetic sublevels to the $F=2$ manifold, where they are pushed out of the trap with light pulses resonant on the S$_{1/2} (F=2)$ to P$_{3/2} (F'=3)$ transition. The magnetic field is then changed adiabatically to the desired value for the spin evolution experiment.
The measured mean trap frequency of the crossed dipole trap is $\approx$490 Hz for thermal atoms, while after evaporating to a pure BEC it is $\approx$140 Hz. The typical density of the thermal sample is $\bar n = 5\,\times\,10^{12}$ cm$^{-3}$. 
The temperature 
is about 100 nK above the critical temperature for BEC in our system. 

Once atoms are prepared in either the pure $m_F=0$ or $m_F=-1$ state, we transfer the atoms to a coherent superposition spin state using a short radio frequency (RF) pulse. 
These states evolve differently 
during the resonant RF pulse. 
In both cases the population distribution in the final superposition state is
$\rho_0\approx0.5$ and $\rho_{+1}=\rho_{-1} \approx 0.25$ but the relative phase $\theta$ 
is different. Cold (or BEC) atoms that are prepared  in the $m_F=0$ state have an initial relative phase of $\theta$ = $\pi$, whereas when the atoms  start in the $m_F=-1$ sublevel the prepared phase is $\theta$ =~0.  These phases result in different behavior for the amplitude and frequency of the population oscillations \cite{You}, as well as the damping. 

After the RF pulse the initial superposition is allowed to evolve for a variable amount of time, after which the population in the $m_F$=0 state is detected by adiabatically sweeping a microwave field over the $F=1$ to $F=2$ transition, and detecting the $F=2$ state atoms by absorption imaging on the $F=2$ to $F'=3$ transition.   
We observe clear oscillations in the $m_F=0$ spin population for the thermal gas for different magnetic fields 
and initial phases.  An example is shown in Fig.~1(b).  The  oscillations are damped on the time scales shown in Figs.~2(b) and 3(b).

We make sinusoidal fits to the oscillations 
and extract the amplitudes and frequencies as functions of the magnetic field. 
These data are then fit using the theory of Ref. \cite{You}, replacing $c_2$ with $2c_2$ for our thermal gas,  and using the total atom number and initial population fraction in $m_F=0$, which was measured separately.   The parameter $c_2\bar n$ is found from the fit.  The data and fits are shown in Figs.~2 and 3, and the fits are consistent with the data within the uncertainties, which are one standard deviation from the sinusoidal fits.

\begin{figure}[h!]
\centering
\includegraphics[width=1\linewidth]{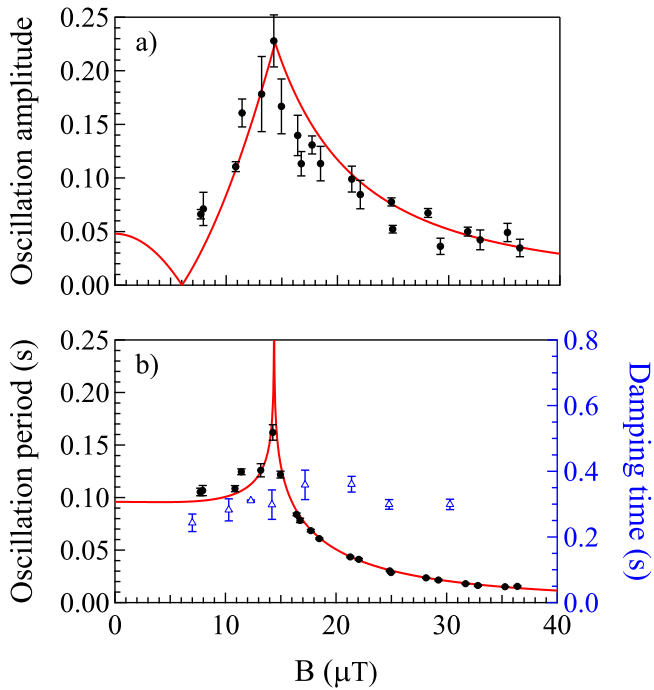} 
 \caption{\label{minus1} (Color online) (a) Amplitude, (b) period (black points), and damping time (blue triangles) of the spin-population oscillations of thermal atoms that are prepared in the $m_F=-1$ state 
(initial $\theta = 0$)
as functions of the magnetic field. The solid curves are predictions from the single-mode theory (see text) with the initial fraction in $m_F = 0$ set to the measured mean of $\rho_0 = 0.45$. 
The theory curve would go to zero at zero field in panel (a) if $\rho_0 = 0.5$.  } 
\end{figure}

\begin{figure}[h!]
\centering
\includegraphics[width=1\linewidth]{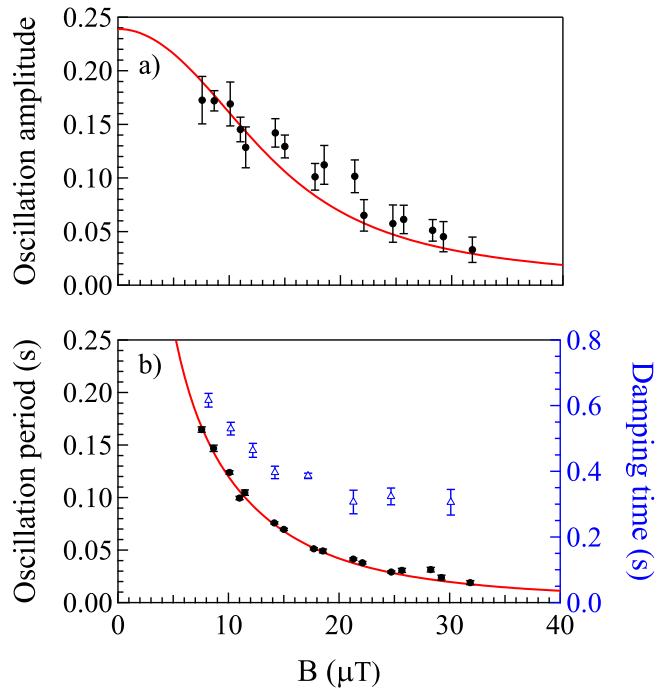} 
\caption{\label{zero} (Color online) (a) Amplitude, (b) period (black points), and damping time (blue triangles) of spin-population oscillations of thermal atoms that are 
prepared in the $m_F=0$ state 
(initial $\theta = \pi$)
as functions of the magnetic field.  Solid curves are predictions of the amplitude and period from the single-mode theory (see text) with the initial $m_F = 0$ fraction set to the measured mean $\rho_0 = 0.52$.} 
\end{figure}

For an initial $\theta=0$ the oscillation amplitude peaks and the period diverges. 
This happens when the difference in collisional energies matches the difference in quadratic Zeeman energies ($q = 2 c_2 \overline{n}$) for the oscillating states. This occurs at $\approx$14.5$~\mu$T, as shown in Fig.~2(a). 
Because of the sharp, resonant behavior the fit is sensitive to the value of $c_2$. From the fit in Fig.~2, we obtain $c_2\bar{n}/h$ = 2.62(2) Hz with the measured density of $\bar{n}=4.9(2)\times10^{12}$ cm$^{-3}$. We find $c_2/h=5.3(2)\times10^{-13}$ Hz cm$^3$, which agrees within the error with the theoretical value of $c_2/h = 5.5(4)\times10^{-13}$ Hz cm$^3$ \cite{Eite}. The uncertainty is one standard deviation and only reflects statistical uncertainty.

A thermal gas with an initial phase of $\theta=\pi$, 
shown in Fig.~3, exhibits quite different behavior.
The amplitude has a maximum at zero magnetic field and goes to zero at large fields.  The period diverges at zero field.  Above 40~$\mu$T the oscillation amplitude cannot be discerned above the experimental noise level. 
Fitting the data we find $c_2 = 5.4(8) \times10^{-13}$ Hz cm$^3$, consistent with the result from Fig.~2, but with a much larger uncertainty.

We have followed the population oscillations in some cases over tens of periods. At longer times we observe amplitude damping of the oscillations, as well as a change of the mean number of atoms in the $m = 0$ state.
The total atom number is observed to be constant over the time scale of the measurements.
In Figs.~2(b) and 3(b) we plot, in addition to the oscillation period, the damping time of the oscillation amplitude, obtained by fitting to a damped sine wave plus a decaying baseline.
Uncertainties in the damping times are one standard deviation of the mean, from 3 - 5 measurements.  The damping times are approximately equal for $\theta = 0$ or $\pi$ above 15 $\mu$T, but diverge at low fields and differ by more than a factor of two below 10  $\mu$T.
Although we do not understand the damping mechanism at this time, it is perhaps not surprising that it should vary, given that the spin dynamics vary dramatically with $\theta$ here as well.  
A theoretical model of spin damping in Ref.~\cite{Endo2008} predicts a time scale proportional to $(c_2)^2$ that is one order of magnitude slower than we observe. This discrepancy might be a limitation of the linearization around thermal equilibrium of the spatial distribution in the collisional integral and contributions from terms proportional to $c_0 c_2$ and/or $(c_0)^2$ might be responsible for the change. It is also possible that uncontrolled experimental issues such as field gradients contribute.

In conclusion, we 
report the first observations of coherent spin-mixing collisions in a thermal spin-1 Bose gas.  We show that this multi-mode thermal gas obeys a theoretical model that displays the same separation of the spatial and spin degrees of freedom  as for a BEC in the single-mode regime.  The crucial difference is that the interaction strength $c_2$ is replaced by $2c_2$ for the thermal case.
The measured population oscillations agree well with theory. Fits to the data return a value of $c_2/h$ in good agreement with that of previous predictions \cite{Eite}. In addition, the damping of the oscillations is found to vary with the initial conditions and magnetic field, but to be an order of magnitude faster than predicted.  
We are currently investigating spin-1 BEC/thermal-atom mixtures, which promise to reveal information about spin-locking between the thermal and Bose-condensed components, as seen in \cite{Lewandowski2}. Finally,  while Heisenberg-limited interferometry with non-classical spin-matter-waves using BECs has been suggested \cite{Klempt}, 
our results suggest that such experiments based on the spin degree of freedom \cite{Marino} could also be performed with a cold thermal gas.

\begin{acknowledgments}
We thank Ryan Wilson for providing useful insight on the theory of the damping of the spin coherences. PFG was supported by a RSE/Scottish Government Marie Curie Personal Research Fellowship. ET acknowledges support from ARO.  JPW acknowledges support from the NRC Postdoctoral Research Associateship program.  RB acknowledges support from the JQI.
\end{acknowledgments}


\begin{thebibliography}{39}
\bibitem{footie}
In deriving Eq.~(6), we have used the fact that the Hartree and Fock terms provide identical contributions. (This follows from the identity that, for any 3$\times$3 matrix $A$, $ S_{\alpha} A S_{\alpha} = {\rm Tr}(A) - A + {\rm Tr} (S_{\alpha} A) S_{\alpha}$.)
\end{thebibliography}
\end{document}